\documentclass[conference]{IEEEtran}
\IEEEoverridecommandlockouts
\usepackage{cite}
\usepackage{amsmath,amssymb,amsfonts}
\usepackage{graphicx}
\usepackage{textcomp}
\usepackage{xcolor}
\usepackage{float}
\usepackage{hyperref}

\usepackage[ruled,linesnumbered]{algorithm2e}
\usepackage{bm} % For bold math symbols
\usepackage{varwidth} % For variable width boxes

\def\BibTeX{{\rm B\kern-.05em{\sc i\kern-.025em b}\kern-.08em
    T\kern-.1667em\lower.7ex\hbox{E}\kern-.125emX}}

\makeatletter
\let\old@ps@headings\ps@headings
\let\old@ps@IEEEtitlepagestyle\ps@IEEEtitlepagestyle
\def\confheader#1{%
\def\ps@IEEEtitlepagestyle{
\old@ps@IEEEtitlepagestyle
\def\@oddhead{\strut\hfill#1\hfill\strut}
\def\@evenhead{\strut\hfill#1\hfill\strut}
}
\ps@headings
}
\makeatother
\confheader{
\small{Proceedings of the 11th RSI International Conference on Robotics and Mechatronics (ICRoM 2023), Nov. 19-21, 2022, Tehran, Iran} 
}
\usepackage[pscoord]{eso-pic}
\newcommand{\placetextbox}[3]{
\setbox0=\hbox{#3}
\AddToShipoutPictureFG*{ \put(\LenToUnit{#1\paperwidth},\LenToUnit{#2\paperheight}){\vtop{{\null}\makebox[0pt][c]{#3}}}
}
}
\placetextbox{.23}{0.055}{\textbf{\small{979-8-3503-0810-5/23/\$31.00~\copyright 2023 IEEE}}}

\begin{document}

\title{Autonomous Driving using Residual Sensor Fusion and Deep Reinforcement Learning\\

}

\author{\IEEEauthorblockN{Amin Jalal Aghdasian, Amirhossein Heydarian Ardakani, Kianoush Aqabakee, Farzaneh Abdollahi}

\IEEEauthorblockA{
amin.aghdasian@aut.ac.ir, ah.heydarian@aut.ac.ir, kianoush.aqabakee@aut.ac.ir, f\_abdollahi@aut.ac.ir}
\\
\IEEEauthorblockA{
\textit{Department of Electrical Engineering} \\
\textit{Amirkabir University of Technology (Tehran Polytechnic)}\\
Tehran, Iran}
}

\maketitle

\begin{abstract}
This paper proposes a novel approach by integrating sensor fusion with deep reinforcement learning, specifically the Soft Actor-Critic (SAC) algorithm, to develop an optimal control policy for self-driving cars. Our system employs a two-branch fusion method for vehicle image and tracking sensor data, leveraging the strengths of residual structures and identity mapping to enhance agent training. Through comprehensive comparisons, we demonstrate the efficacy of information fusion and establish the superiority of our selected algorithm over alternative approaches. Our work advances the field of autonomous driving and demonstrates the potential of reinforcement learning in enabling intelligent vehicle decision-making.
\end{abstract}


\section{Introduction}

%\textbf{Autonomous driving}

In recent years, Autonomous Driving (AD) has become a tangible reality, and as this technology progresses, it is imperative to ensure the ongoing safety and reliability of this transportation mode. With the potential to significantly decrease accidents stemming from human error, AD emerges as a promising solution for enhancing road safety. According to data from the National Highway Traffic Safety Administration (NHTSA), a staggering 94\% of all traffic accidents can be attributed to human error or its associated factors. AD systems come equipped with advanced safety features like collision avoidance systems and sophisticated sensors capable of preemptively averting accidents. These safety mechanisms consistently monitor the vehicle surroundings and can react much more swiftly than a human driver when confronted with unexpected situations~\cite{a2}. 

%\textbf{Control strategies}

AD relies on a variety of control strategies, including classic controllers such as PID~\cite{b1}, MPC~\cite{b2}, LQR~\cite{b3}, and Artificial Intelligence (AI) controllers such as imitation learning based control~\cite{b4} and RL based control. PID control is simple but requires tuning, while MPC handles complex scenarios. LQR control optimizes a cost function efficiently, but it requires accurate knowledge of system dynamics and may need to be redesigned for significant changes~\cite{b6}. Model-free approaches offer distinct advantages over model-based methods in various contexts. They shine in complex and dynamic environments where precise modeling is challenging, providing adaptability to uncertain or evolving dynamics. Their ease of implementation and reduced need for manual tuning make them appealing for practical applications. Model-free reinforcement learning, in particular, stands out for its ability to continuously learn and adapt, achieving impressive performance across diverse domains. These methods are well-suited for scenarios where modeling is impractical or resource-intensive and excel in handling novel or unforeseen situations without the constraints of fixed assumptions. Imitation learning utilizes expert demonstrations but may encounter errors when faced with unseen events or situations. In contrast, RL control offers adaptability, continuous learning, and the ability to handle diverse conditions~\cite{b9}.

%\textbf{Reinforcement Learning}

Deep reinforcement learning (DRL) represents a subset of machine learning methodologies in which an agent acquires decision-making capabilities through interactions with its environment~\cite{q1}. Within AD, RL has found applications in developing driving policies~\cite{q2}. These algorithms have successfully tackled complex problems such as Markov Decision Problems (MDPs). However, as highlighted earlier, it can be challenging for the algorithm to learn from all possible states and determine the optimal driving path. ~\cite{b10} employs a Deep Q-Network (DQN) and Long-Short-Term Memory (LSTM) architecture, along with a novel observation input, to enable autonomous agents to navigate intricate scenarios autonomously and~\cite{b9} employs DQN and Deep Deterministic Policy Gradient (DDPG) algorithms to train autonomous vehicle control models within a realistic simulator. DQN and DDPG achieve the desired control objectives, with DDPG demonstrating superior performance since it is suitable for continuous tasks. In~\cite{b11}, It was developed two approaches: pre-training Soft Actor-Critic (SAC) with Learning from Demonstrations (LfD) and an online combination of SAC, LfD, and  Learning from Interventions (LfI). 
Given the recent advances in DRL algorithms for self-driving cars, enhancing their performance can be achieved by incorporating more complex methods for data utilization in the network structures.

%\textbf{Sensor Fusion}
Sensor fusion enables AD systems to perceive and understand the surrounding environment accurately. This technology integrates data from diverse sensors, aligning them to generate different representations of the vehicle's surroundings~\cite{s1,s2,s4}. This fusion of information not only enhances the robustness and reliability of the perception system in various driving scenarios~\cite{s5,s6} but also helps in mitigating the limitations of individual sensors, thus improving the overall performance of autonomous vehicles~\cite{s7,s8}.
In the context of RL, sensor fusion becomes even more crucial. RL-based control systems learn to make decisions by interacting with the environment, and the quality of these decisions depends on perceived information quality~\cite{s9,b9}. With the integrated and comprehensive environmental models provided by sensor fusion, RL algorithms can better understand the state of the environment, resulting in more accurate predictions and safer decision-making~\cite{s11}. Moreover, incorporating multi-modal fusion into DRL frameworks has demonstrated significant potential in enhancing decision-making and reliability in complex environments~\cite{s12,s13}.

%\textbf{Our CONTRIBUTIONS}
This study presents a novel vehicular control methodology that leverages sensor data fusion and reinforcement learning. Our approach incorporates a sensor fusion module consisting of a dual branch architecture, effectively combining information from the vehicle image and tracking sensors. We employed residual blocks within this module to enhance feature extraction and representation learning. By integrating the Soft Actor-Critic (SAC) with the sensor fusion technique, our proposed method unifies the outputs of image sensors with the vehicle speed and position, resulting in an optimal structure for generating control commands and faster convergence.
To evaluate the effectiveness of our algorithm, we conducted comprehensive training and validation procedures using the CARLA simulation platform. This platform offers a realistic environment to assess our algorithm capabilities and simulate real-world driving scenarios. Furthermore, this paper includes an extensive comparative analysis, highlighting the potential advantages of our approach compared to existing state-of-the-art methods in the field.

%\textbf{Paper Structure}
The paper is structured into five sections. Section 1 provides an introduction, offering insights into autonomous driving approaches. Section 2 delves into the background information. Section 3 outlines the proposed methodology, highlighting the essential aspects. The results and discussion section, Section 4, presents the research findings. Finally, Section 5, the conclusion, summarizes the paper and suggests potential future research directions.

\section{Soft Actor-Critic (SAC)~\cite{f1}}

Soft Actor-Critic (SAC) represents an off-policy DRL algorithm that delivers efficient learning while preserving the advantages associated with entropy maximization and stability~\cite{f1,f2,f3}.
SAC bridges between stochastic policy optimization and DDPG~\cite{f4} methods. While not a direct successor to Twin Delayed DDPG~\cite{f5}, it utilizes the clipped double-Q technique and benefits from target policy smoothing due to its inherent stochasticity. SAC emphasizes entropy regularization, wherein the policy seeks a balance between expected return and policy randomness. This approach enhances exploration, aiding later learning stages, and guards against premature convergence to suboptimal solutions.

%\textbf{s2}

To delve into SAC, we must initially introduce the context of entropy-regularized reinforcement learning. In this framework, value functions are defined by slightly different equations. The entropy $H$ is determined as follows:
\begin{equation}
    \label{eq:entropy}
    H(P) = E_{x \sim P}[-\log P(x)]
\end{equation}
Where $x$ is a random variable with a probability mass or density function P. In entropy-regularized reinforcement learning, the agent receives an additional reward at each time step proportionate to the policy entropy for that specific time step. This adjustment transforms the RL goal, which is the reward maximization problem, into:
\begin{equation}
    \label{eq:fir}
    \small
    \pi^* = \arg \max_{\pi} \left[E_{\tau \sim \pi} \left[\sum_{t=0}^{\infty} \gamma^t \left(R(s_t, a_t, s_{t+1}) + \alpha H(\pi(\cdot \mid s_t))\right)\right]\right]
\end{equation}

In~\eqref{eq:fir}, a transition is denoted by $(s_t, a_t, r_t, s_{t+1}, a_{t+1})$, including state, action, and reward.
$\alpha $ represents the trade-off coefficient, $\gamma$ is discount factor and $\pi(\cdot \mid s_t)$ is policy in state $s_t$ . We consider an infinite-horizon discounted. Under this assumption, we introduce modified value functions. Specifically, the action-state value function $Q^{\pi}$ is adapted to incorporate entropy bonuses at each time step:

\begin{equation}
    \label{eq:Q_function}
    \begin{split}
    \resizebox{0.9\hsize}{!}{$
    Q^{\pi}(s_t,a_t) = E \left[R(s_t,a_t,s_{t+1}) + \gamma(Q^{\pi}(s_{t+1},a_{t+1}) + \alpha H(\pi(\cdot \mid s_{t+1})))\right]$} \\
    \resizebox{0.9\hsize}{!}{$
    =E \left[R(s_t,a_t,s_{t+1}) + \gamma(Q^{\pi}(s_{t+1}a_{t+1}) -\alpha \log (\pi(a_{t+1} \mid s_{t+1})))\right]$}
    \end{split}
\end{equation}
\vspace{0.2pt}

This approach considers two parameterized soft Q-functions, $Q(s_t, a_t)$, and a policy represented as $\pi(a_t|s_t)$. Each component relies on specific parameter sets, denoted as $\phi_1$, $\phi_2$, and $\theta$, respectively. For instance, the Q-functions can be modeled as expressive neural networks.
Actions are resampled as $\tilde{a}_{t+1} \sim \pi (\cdot  \mid s_{t+1})$ under the current policy instead of being drawn from the replay buffer. Therefor, it is essential to distinguish newly sampled actions. The parameters of the soft Q-function can then undergo training aimed at minimizing the soft Bellman residual:
\begin{equation}
    \label{eq:JQ_function}
    J_{Q_i}(\theta) = E_{(s_t,a_t) \sim D} \left[\frac{1}{2} \left(Q_\phi(s_t,a_t) - \hat{Q}(s_t,\tilde{a}_t)\right)^2\right]
\end{equation}
where the target $Q$ is represented as:
\begin{multline}
    \label{eq:hat_Q_function}
    \small
    \hat{Q}(s_t,\tilde{a}_t) = r_t +  \gamma \left (Q(s_{t+1},\tilde{a}_{t+1}) - \alpha \log (\tilde{a}_{t+1} \mid s_{t+1}) \right )
\end{multline}
Equation~\eqref{eq:JQ_function} can be optimized using gradient descent:
\begin{equation}
    \label{eq:gradient_JQ}
    \small
    \hat{\nabla}_\phi J_Q(\phi) = \nabla_\phi Q_\phi(s_t,a_t)\left(Q_\phi(s_t,a_t) - \hat{Q}(s_t,a_t)\right)
\end{equation}

The policy parameters can be acquired by minimizing the Kullback-Leibler divergence~\cite{f1}. By reparameterizing the expectation, we can now define the policy objective as:

\begin{equation}
    \label{eq:Jpi_function_rewrite}
    \small
    J_\pi (\theta) = E_{s_t \sim D} \left[ \log \pi_\theta (\tilde{a}_\theta (s_t)\mid s_t) - Q_\phi(s_t,\tilde{a}_\theta (s_t))\right]
\end{equation}

Then, we will approximate the gradient of~\eqref{eq:gradient_Jpi} as follows:
\begin{multline}
    \label{eq:gradient_Jpi}
    \small
    \hat{\nabla}_\theta J_\pi(\theta) = \nabla_\theta \log(\pi_\theta(\tilde{a}_t \mid s_t)) 
    \\ + \left(\nabla_{\tilde{a}_t} \log(\pi_\theta(\tilde{a}_t \mid s_t)) - \nabla_{\tilde{a}_t} Q(s_t,\tilde{a}_t)\right) \nabla_\theta \tilde{a}_\theta(s_t)
\end{multline}

Through iterative interactions and data collection, the Q-function and policy networks will reach a state of convergence, which enables the agent to obtain the maximum reward in each episode.

\section{Methodology}

\subsection{Problem Formulation}

The autonomous driving problem can be delineated into several primary objectives. These objectives include arriving at the designated destination while ensuring collision avoidance and navigation within the road lane with minimal error.
Each state $s_t$ encompasses the data acquired from the car front camera and the tracking sensor. These inputs are combined in a structured manner, as described in the subsequent section, to generate the desired control command. Car navigation relies on two controls: longitudinal and lateral. Consequently, the DRL agent produces two control commands: acceleration and steering angle. The car acceleration range is [0,1] and the steering range is [-1,1]. Furthermore, the episode return $R_t$ for the problem is established by~\eqref{eq:return}, ensuring the car navigates between the designated boundaries towards the specified destination.

\begin{equation}
    \label{eq:return}
    R_t = \sum_{t}^{}\left | v_t cos(\phi_t) \right |-\left | v_t sin(\phi_t) \right |-\left | v_t \right |\left | d_t \right |
\end{equation}

\begin{figure}[h]
    \centering
    \includegraphics[width=0.6\linewidth]{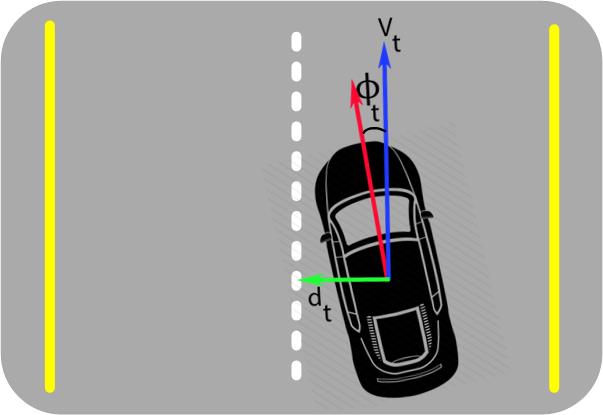}
    \caption{Reward function parameters based on vehicle orientation.}
    \label{fig:car}
\end{figure}

Where the angle $\phi_t$ represents car orientation relative to the road, the cosine term denotes the speed aligned with the road length, and the sine term represents the speed perpendicular to the road. By formulating the return function as~\eqref{eq:return}, the RL agent maximizes car velocity $v_t$ by increasing the first term and decreasing the second one. Furthermore, the third term aims to minimize vehicle deviation from the center of the lane, as indicated by the distance $d_t$. 
Fig. \ref{fig:car} visually represents the reward parameters of the vehicle on the road.

The episode terminates in the event of a collision or the car deviating from its designated lane, resulting in a reward of $-200$. Conversely, upon reaching the goal destination, a reward of $100$ is recorded for the episode. Consequently, the agent aims to navigate to the destination without collisions to maximize the overall return.

\subsection{Sensor Data Fusion}

\begin{figure}[h]
    \centering
    \includegraphics[width=0.95\linewidth]{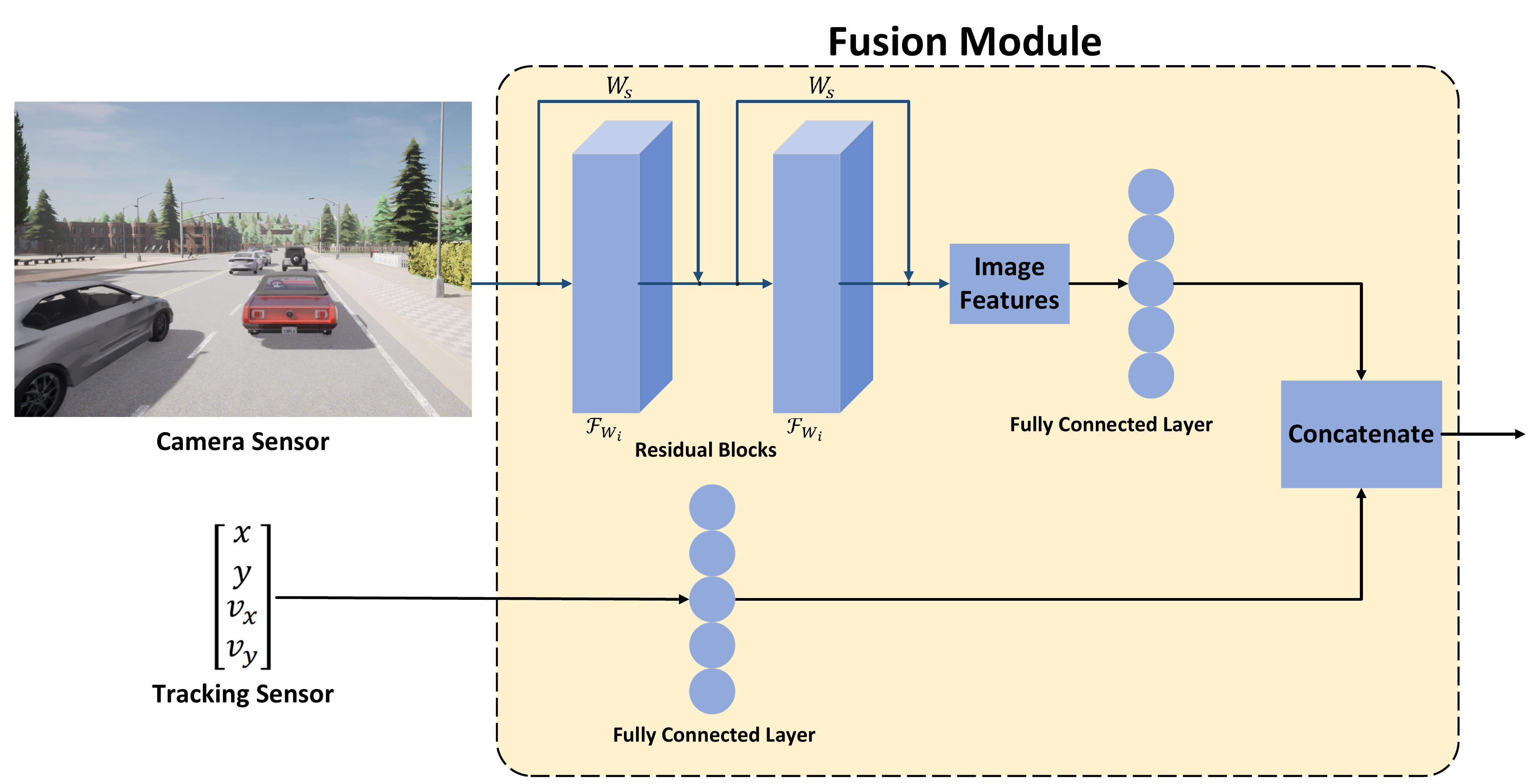}
    \caption{Sensor fusion module.}
    \label{fig:fusion}
\end{figure}

The fusion module within our RL agent enhances its decision-making capabilities and overall performance by integrating various data sources and unifying information. 
This module plays a vital role in both the actor and critic components, effectively improving car navigation by identifying the optimal approach to information fusion. A visual representation of the fusion module structure is presented in Fig.~\ref{fig:fusion}.

The fusion module within the system employs a dual-branch approach for integrating the information derived from the image and the tracking sensors. The image branch encompasses multiple residual blocks and a fully connected layer. 
The residual block extracts image features using multiple convolutional layers $\mathcal{F}_{W_i}$ and a shortcut path~\eqref{eq:fusion}. In order to ensure compatibility of dimensions, a projection $W_s$ is additionally employed using 
a convolution layer in the direction of the shortcut path. Incorporating skip connections within these residual blocks facilitates the direct propagation of information, thereby mitigating the vanishing gradient problem~\cite{m1}. Moreover, a fully connected layer transforms the features extracted from the image into the desired dimensional representation. 

\begin{equation}
    \label{eq:fusion}
    features = \mathcal{F}_{W_i}(image) + W_simage
\end{equation}

In the other branch, the data obtained from the tracking sensor undergoes resizing via a fully connected layer. The fused information is transmitted throughout the remaining network structure by establishing a connection between these two branches.

\subsection{Proposed Method Architecture}
Our proposed method represents a combination of sensor fusion and SAC architecture for controlling autonomous vehicles. SAC is chosen in this research due to its vital attributes in stability, exploration, and robustness~\cite{f1}. The fusion structure proposed in the previous section is implemented at the outset of each function approximator (Q-network 1, Q-network 2, and the actor). This involves concatenating camera and tracking sensor data features, utilizing them as input to the model while carefully considering the importance of each data component in each function approximator to achieve optimal results. 

\begin{figure}[h]
    \centering
    \includegraphics[width=0.95\linewidth]{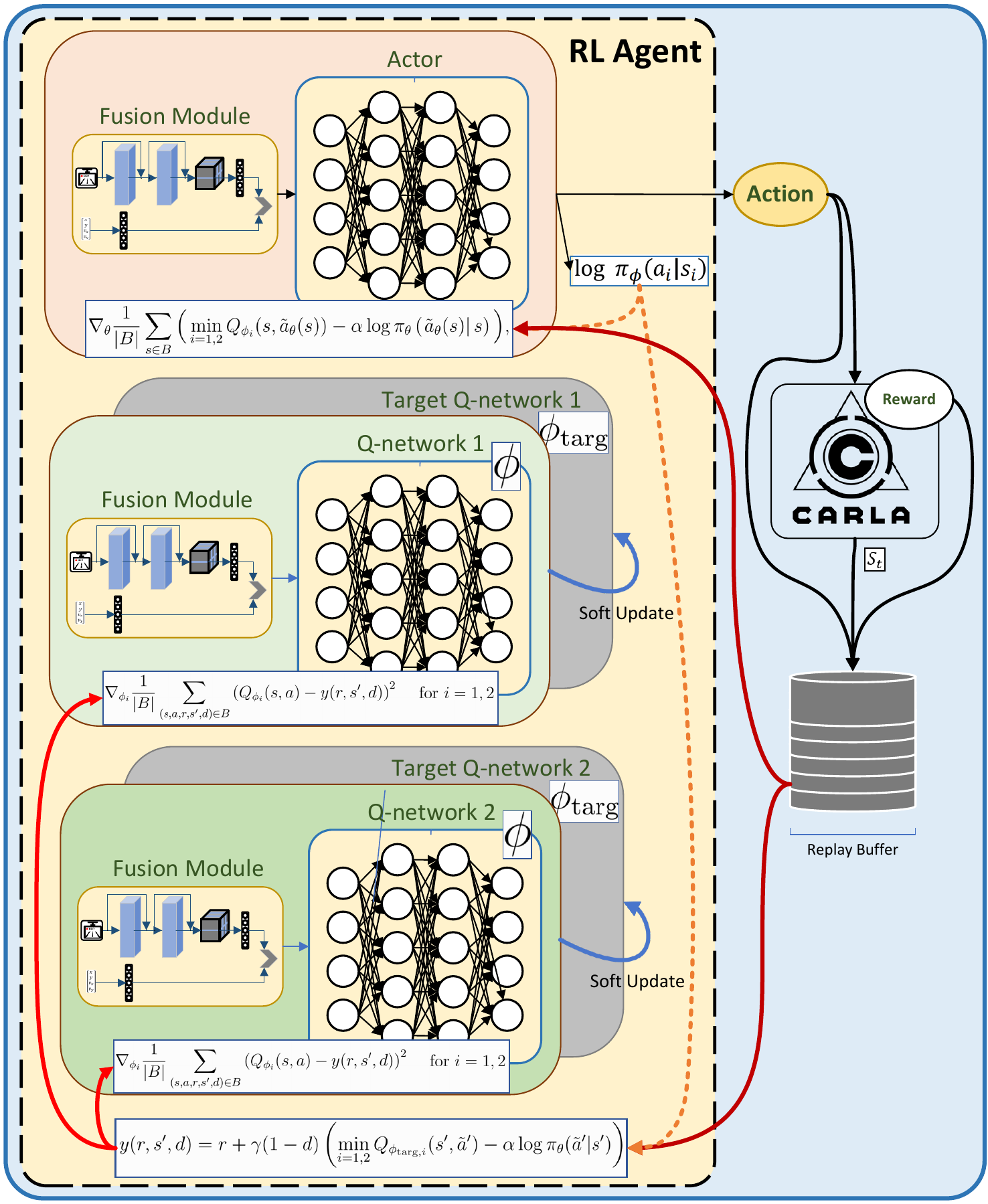}
    \caption{Proposed DRL control scheme}
    \label{fig:proposed}
\end{figure}

As illustrated in Fig.~\ref{fig:proposed}, the actor model is trained based on the approximated action-value function, and after a sufficient number of training episodes, we can attain the desired level of performance. The reason for merging multiple inputs with the proposed fusion structure is to extract features at the earlier layers and combine these representations at the subsequent layers to create a more comprehensive understanding of the environment. Having a greater variety of sensor types implies increased data dimensions collected from the environment. Higher dimensionality provides a richer source of information for making ultimate decisions. Nevertheless, selecting the most concise data representation that exhibits the highest correlation with the desired output remains critical. We used data concatenation and mapped it to a reduced-dimensional space in subsequent layers with mentioned specifications. However, this approach presents some challenges. We encounter issues such as vanishing gradients and slow convergence during feature extraction and selection processes. To navigate these challenges effectively, we added the residual structures and SAC algorithm.
The pseudocode of the proposed approach is shown in the Alg.~\ref{alg:SAC}.

\begin{algorithm}[h]
    \caption{Pseudocode of the proposed approach}
    \label{alg:SAC}
    \begin{varwidth}{\linewidth}
    \textbf{Input:} Initialize networks parameters $\bm{\theta}$, $\bm{\phi_1}$, $\bm{\phi_2}$ and  replay buffer $\mathcal{D}$
    \end{varwidth}

    \textnormal{Initialize}  target networks parameters $\bm{\phi_{\text{target 1}}} \leftarrow \bm{\phi_1}$, $\bm{\phi_{\text{target 2}}} \leftarrow \bm{\phi_2}$

    \For{each episode}{
        \For{each CARLA step}{
            $a_t \leftarrow \pi_{\theta}(a_t | s_t)$\\
            $s_{t+1} \leftarrow p(s_{t+1} | s_t, a_t)$\\
            $D \leftarrow D \cup \{(s_t, a_t, r(s_t, a_t), s_{t+1})\}$\\
            Reset environment$\leftarrow$ if ${s_{t+1}}$ is terminal state
        }
        \For{each gradient step}{
             Randomly sample batch from transitions $\mathcal{D}$\\
             Compute targets for Q-functions:\\
             $y(r,s') = r + \gamma \min_{i=1,2} \left[Q_{\phi_{\text{target},i}}(s',{\tilde{a}}') - \alpha \log(\pi_\theta({\tilde{a}}' | s^{'}))\right]$ \\
             Update Q-functions by one step of gradient descent for $i=1,2$:\\
             $\nabla_{\phi_i} \frac{1}{|B|} \sum_{\left ( s,a,r,s{}' \right )\epsilon B}^{} \left(Q_{\phi_i}(s,a) - y(r,s')\right)^2$ \\
             Update policy by one step of gradient ascent:\\
             $\nabla_{\theta} \frac{1}{|B|} \sum_{s\epsilon B}^{} (\min_{i=1,2}Q_{\phi_i}(s,\tilde{a}_{\theta }(s))$ \\
             $\quad \quad \quad \quad \quad \quad \quad \quad \quad \quad - \alpha \log(\pi_\theta(\tilde{a}_{\theta }(s) \mid s)))$ \\
             Update target networks:\\
             $\phi_{target, i}\leftarrow \rho \phi _{target, i}+\left (1-\rho  \right )\phi_{i} \ for \ i=1,2$ \\
        }
    }
\end{algorithm}

\section{Results and Discussion}

We selected CARLA~\cite{q8}, an open-source simulator designed for AD research, for this study. CARLA provides a Python library that facilitates interaction with the environment, enabling dynamic modifications of the spawning and removal of cars and objects within the simulated environment. To assess the efficacy of the proposed method against alternative algorithms, SAC algorithm was subjected to three distinct training scenarios in CARLA. These scenarios encompassed using the sensor fusion block, solely relying on camera imagery, and tracking sensor data under similar situations. Furthermore, we employed the DDPG algorithm used in previous studies, following a similar training procedure to establish the system superiority compared to previous works~\cite{b9}.

\subsection{Training}

We designed three different training configurations to assess the impact of sensor fusion on the RL agent performance. In the first configuration, we utilized sensor fusion, combining both image and tracking sensor data, to inform the agent decision-making process. In the second configuration, we isolated the use of only image data, while in the third configuration, we relied solely on tracking sensor data for decision-making. These configurations were designed to highlight the versatility and adaptability of our algorithm under varying data input scenarios.
The same training process has been repeated for the DDPG algorithm to compare the performance of the proposed algorithm with previous works~\cite{b9}.

\begin{figure}[h]
    \centering
    \includegraphics[width=1\linewidth]{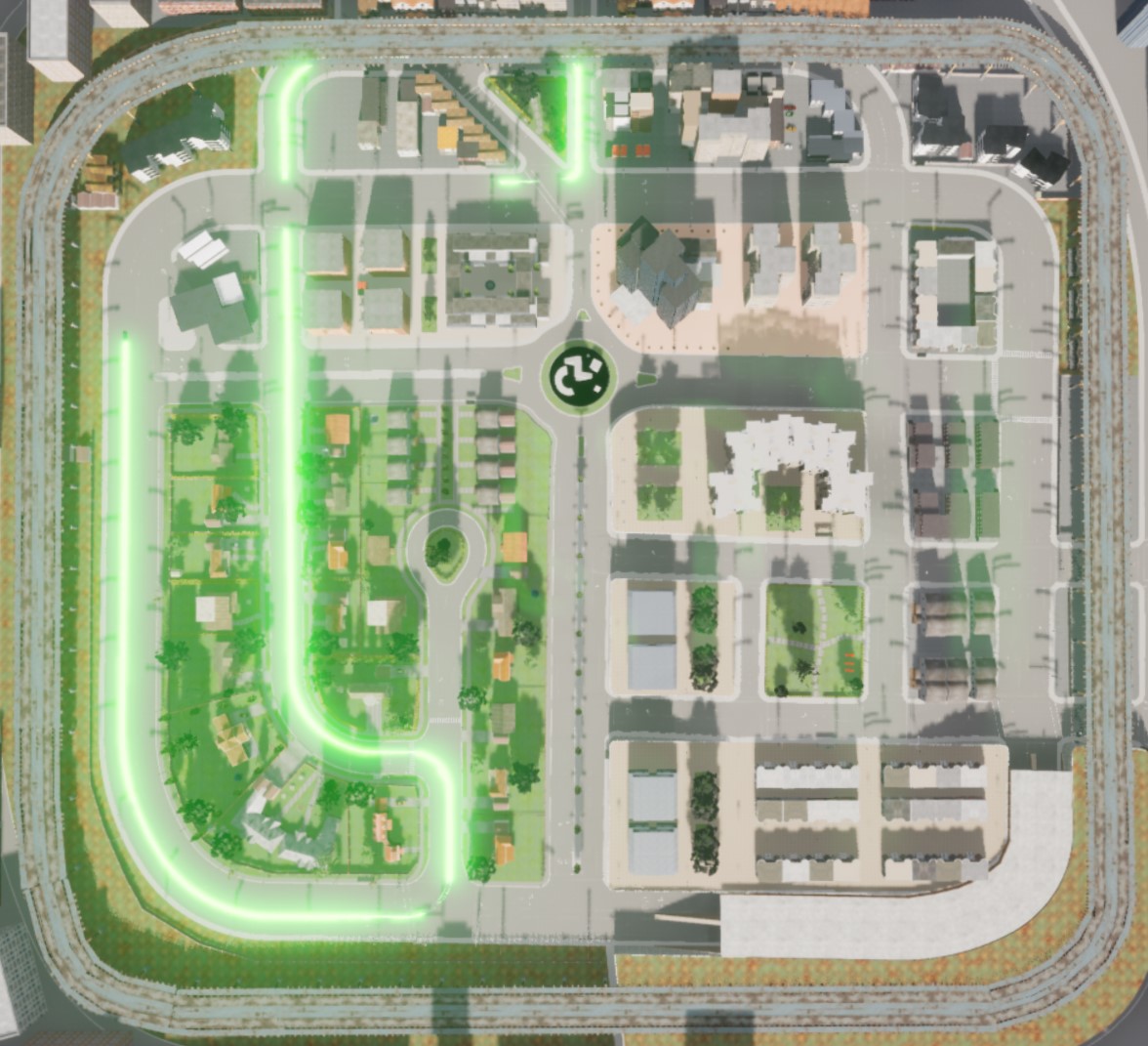}
    \caption{Random path generation in CARLA simulator.}
    \label{fig2}
\end{figure}

To create episodic tasks for training, we randomly choose two points on the town roadway and create a path from the start point to the end point. 
Fig.~\ref{fig2} shows an example of a random path generated on the map.
By training on different town maps and weather conditions, our DRL agent will become more general and robust, capable of handling various driving tasks and uncertainties.

The camera image is mapped to 100 features using the proposed residual structure. Then, a fully connected layer maps 16 features from the tracking sensor to be concatenated with image features. In Q-networks, we have 116 input features for our agent; by adding two actions, throttle and steering wheel angle, 118 features will be delivered. We trained the agent for 5000 episodes with Adam optimizer and a learning rate value of 0.0001. Also, the buffer size is considered 1e6 to cover a wide range of explorations. The average reward is calculated in each episode, as mentioned before.
The training and testing of the proposed approach were performed on a hardware with a Core i9 5 GHz CPU, Nvidia Geforce RTX 3080 Ti GPU with 32 GB of RAM. 

\begin{figure}[h]
    \centering
    \includegraphics[width=1\linewidth]{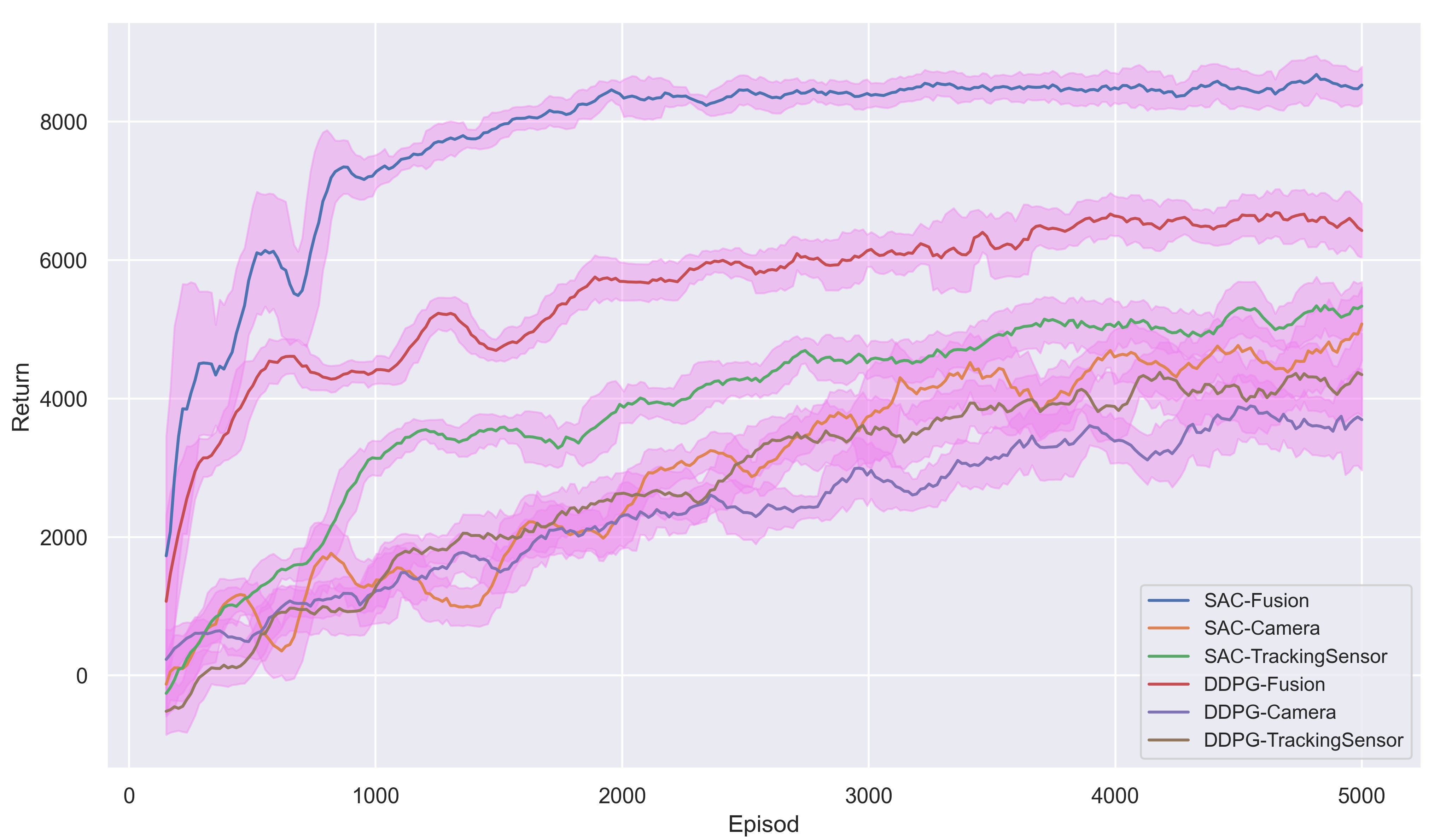}
    \caption{Smoothed plot of average reward per episode for six training configurations.}
    \label{fig:reward}
\end{figure}

As shown in Fig.~\ref{fig:reward}, the proposed fusion structure leads to faster convergence and higher average reward with minor deviation in both SAC and DDPG algorithms compared to using only one sensor data. Also, SAC can unleash the potential of the proposed structure better than DDPG due to the use of the entropy function to enhance optimization and better exploration strategy.

\subsection{Evaluation}

The validation for proposed approaches employs the Root Mean Square Error (RMSE) metric, designed to evaluate the error between each algorithm and the ground truth provided by the CARLA Simulator. This approach ensures that the performance analysis of algorithms adheres to a uniform criterion. 
In order to compare the trained agents, 25 different routes are designated on the map. The agents are navigated along these routes, facilitating the calculation of the RMSE between the actual route and an ideal route derived from the interpolation of the waypoints.
Table~\ref{tab1} shows the RMSE yielded when the agent navigates the route for 25 iterations. Additionally, the table reports the maximum and minimum error recorded along the route, as well as the error standard deviation (std) of the agent driving from the initial to the terminal point.

\begin{table}[h]
\caption{calculated root mean square error in evaluation phase}
\begin{center}
\begin{tabular}{|c|cccc|c|}
\hline
\textbf{Method} & \textbf{\textit{Mean}}& \textbf{\textit{Min}}& \textbf{\textit{Max}}& \textbf{\textit{std}} \\
\hline
%SAC Fusion-based& 0.1$^{\mathrm{a}}$& 0.08& 0.12&0.95  \\
SAC Fusion-based& 0.039& 0.006& 0.079&0.011  \\
SAC Image-based& 0.591& 0.189& 1.073&0.141  \\
SAC Sensor-based& 0.526& 0.175& 0.919&0.12  \\
\hline
DDPG Fusion-based& 0.119& 0.08& 0.158&0.012  \\
DDPG Image-based& 0.752& 0.24& 1.379&0.173  \\
DDPG Sensor-based& 0.678& 0.318& 1.055&0.135  \\
\hline
%\multicolumn{4}{l}{$^{\mathrm{a}}$Sample of a Table footnote.}
\end{tabular}
\label{tab1}
\end{center}
\end{table}

Our evaluation results demonstrate the significant improvement achieved through sensor fusion. We observed a remarkable improvement in the overall error when employing sensor fusion, indicating that the agent made more informed and successful decisions. 
On the other hand, employing solely the image as the agent input shows superior performance compared to relying on the tracking sensor. This disparity arises from the agent ability to determine road segments and anticipate forthcoming actions, particularly during turning maneuvers. Conversely, the information derived from the tracking sensor only provides the agent with the present state of the vehicle.
A comparison between SAC and DDPG also shows the relatively lower error shown by SAC agent. This advantage comes from the effective employment of a stochastic policy, careful consideration of entropy function, and an improved exploration process, all aimed at attaining the optimal policy.
These findings underscore the crucial role of sensor fusion in enhancing the overall performance and robustness of our RL algorithm, making it a valuable approach for a wide range of applications where both image and tracking sensor data are available.

%\subsection{Comparison}

\section{Conclusion}

In this study, we have introduced an intelligent approach for car navigation by integrating sensor fusion into deep reinforcement learning.
To enhance the agent learning performance and ascertain an optimal information integration method, we developed a fusion block that merges data acquired from the vehicle camera and tracking sensor.
Our proposed methodology was evaluated within the CARLA simulation platform, which provides realistic simulations and offers high diversity and repeatability in autonomous driving.

To substantiate the efficacy of our method in contrast to prior researches, we conducted a comprehensive evaluation across three distinct scenarios, encompassing tracking sensor data, image data, and the fusion of both. The results obtained from these evaluations underscore the significant impact of employing sensor fusion along with SAC algorithm to enhance the performance of DRL algorithms in car navigation.
Future works for this research include the real-world implementation of the algorithm, addressing uncertainties inherent in practical settings, and increasing the variety of sensors employed in the navigation system. These explorations hold the promise of advancing the state-of-the-art in autonomous vehicle navigation.

%%%%%%%%%%%%%%%%%%% REFRENCES %%%%%%%%%%%%%%%%%%%

\bibliographystyle{ieeetr}
\bibliography{references}

\begin{thebibliography}{10}

\bibitem{a2}
O.~Zheng, M.~Abdel-Aty, Z.~Wang, S.~Ding, D.~Wang, and Y.~Huang, ``Avoid:
  Autonomous vehicle operation incident dataset across the globe,'' {\em arXiv
  preprint arXiv:2303.12889}, 2023.

\bibitem{b1}
W.~Farag, ``Complex trajectory tracking using pid control for autonomous
  driving,'' {\em International Journal of Intelligent Transportation Systems
  Research}, vol.~18, no.~2, pp.~356--366, 2020.

\bibitem{b2}
K.~Honda, H.~Okuda, T.~Suzuki, and A.~Ito, ``Mpc builder for autonomous drive:
  Automatic generation of mpcs for motion planning and control,'' in {\em 2023
  IEEE Intelligent Vehicles Symposium (IV)}, pp.~1--8, IEEE, 2023.

\bibitem{b3}
J.~Chen, W.~Zhan, and M.~Tomizuka, ``Autonomous driving motion planning with
  constrained iterative lqr,'' {\em IEEE Transactions on Intelligent Vehicles},
  vol.~4, no.~2, pp.~244--254, 2019.

\bibitem{b4}
E.~Bronstein, M.~Palatucci, D.~Notz, B.~White, A.~Kuefler, Y.~Lu, S.~Paul,
  P.~Nikdel, P.~Mougin, H.~Chen, {\em et~al.}, ``Hierarchical model-based
  imitation learning for planning in autonomous driving,'' in {\em 2022
  IEEE/RSJ International Conference on Intelligent Robots and Systems (IROS)},
  pp.~8652--8659, IEEE, 2022.

\bibitem{b6}
J.~H. Yang, W.~Y. Choi, S.-H. Lee, and C.~C. Chung, ``Autonomous lane keeping
  control system based on road lane model using deep convolutional neural
  networks,'' in {\em 2019 IEEE Intelligent Transportation Systems Conference
  (ITSC)}, pp.~3393--3398, IEEE, 2019.

\bibitem{b9}
{\'O}.~P{\'e}rez-Gil, R.~Barea, E.~L{\'o}pez-Guill{\'e}n, L.~M. Bergasa,
  C.~Gomez-Huelamo, R.~Guti{\'e}rrez, and A.~Diaz-Diaz, ``Deep reinforcement
  learning based control for autonomous vehicles in carla,'' {\em Multimedia
  Tools and Applications}, vol.~81, no.~3, pp.~3553--3576, 2022.

\bibitem{q1}
A.~K. Shakya, G.~Pillai, and S.~Chakrabarty, ``Reinforcement learning
  algorithms: A brief survey,'' {\em Expert Systems with Applications},
  vol.~231, p.~120495, 2023.

\bibitem{q2}
K.~B. Naveed, Z.~Qiao, and J.~M. Dolan, ``Trajectory planning for autonomous
  vehicles using hierarchical reinforcement learning,'' in {\em 2021 IEEE
  International Intelligent Transportation Systems Conference (ITSC)},
  pp.~601--606, IEEE, 2021.

\bibitem{b10}
M.~Ahmed, C.~P. Lim, and S.~Nahavandi, ``A deep q-network reinforcement
  learning-based model for autonomous driving,'' in {\em 2021 IEEE
  International Conference on Systems, Man, and Cybernetics (SMC)},
  pp.~739--744, IEEE, 2021.

\bibitem{b11}
M.~Savari and Y.~Choe, ``Online virtual training in soft actor-critic for
  autonomous driving,'' in {\em 2021 International Joint Conference on Neural
  Networks (IJCNN)}, pp.~1--8, IEEE, 2021.

\bibitem{s1}
X.~Zhang, Y.~Gong, J.~Lu, J.~Wu, Z.~Li, D.~Jin, and J.~Li, ``Multi-modal fusion
  technology based on vehicle information: A survey,'' {\em IEEE Transactions
  on Intelligent Vehicles}, vol.~8, no.~6, pp.~3605 -- 3619, 2023.

\bibitem{s2}
J.~Du, X.~Huang, M.~Xing, and T.~Zhang, ``Improved 3d semantic segmentation
  model based on rgb image and lidar point cloud fusion for automantic
  driving,'' {\em International Journal of Automotive Technology}, vol.~24,
  no.~3, pp.~787--797, 2023.

\bibitem{s4}
A.~T. Candan and H.~Kalkan, ``U-net-based rgb and lidar image fusion for road
  segmentation,'' {\em Signal, Image and Video Processing}, vol.~17, pp.~1--7,
  2023.

\bibitem{s5}
S.~Gite, K.~Kotecha, and G.~Ghinea, ``Context--aware assistive driving: an
  overview of techniques for mitigating the risks of driver in real-time
  driving environment,'' {\em International Journal of Pervasive Computing and
  Communications}, vol.~19, no.~3, pp.~325--342, 2023.

\bibitem{s6}
W.~Hou, W.~Li, and P.~Li, ``Fault diagnosis of the autonomous driving
  perception system based on information fusion,'' {\em Sensors}, vol.~23,
  no.~11, p.~5110, 2023.

\bibitem{s7}
H.~Liu, C.~Wu, and H.~Wang, ``Real time object detection using lidar and camera
  fusion for autonomous driving,'' {\em Scientific Reports}, vol.~13, no.~1,
  p.~8056, 2023.

\bibitem{s8}
Q.~Zhang, J.~Liu, and X.~Jiang, ``Lane detection algorithm in curves based on
  multi-sensor fusion,'' {\em Sensors}, vol.~23, no.~12, p.~5751, 2023.

\bibitem{s9}
N.~Chukamphaeng, K.~Pasupa, M.~Antenreiter, and P.~Auer, ``Learning to drive
  with deep reinforcement learning,'' in {\em 2021 13th International
  Conference on Knowledge and Smart Technology (KST)}, pp.~147--152, IEEE,
  2021.

\bibitem{s11}
J.~Dong, S.~Chen, Y.~Li, R.~Du, A.~Steinfeld, and S.~Labi, ``Space-weighted
  information fusion using deep reinforcement learning: The context of tactical
  control of lane-changing autonomous vehicles and connectivity range
  assessment,'' {\em Transportation Research Part C: Emerging Technologies},
  vol.~128, p.~103192, 2021.

\bibitem{s12}
Y.~H. Khalil and H.~T. Mouftah, ``Exploiting multi-modal fusion for urban
  autonomous driving using latent deep reinforcement learning,'' {\em IEEE
  Transactions on Vehicular Technology}, vol.~72, no.~3, pp.~2921--2935, 2022.

\bibitem{s13}
W.~Liu, Z.~Xiang, H.~Fang, K.~Huo, and Z.~Wang, ``A multi-task fusion
  strategy-based decision-making and planning method for autonomous driving
  vehicles,'' {\em Sensors}, vol.~23, no.~16, p.~7021, 2023.

\bibitem{f1}
T.~Haarnoja, A.~Zhou, P.~Abbeel, and S.~Levine, ``Soft actor-critic: Off-policy
  maximum entropy deep reinforcement learning with a stochastic actor,'' in
  {\em International conference on machine learning}, pp.~1861--1870, PMLR,
  2018.

\bibitem{f2}
T.~Haarnoja, A.~Zhou, K.~Hartikainen, G.~Tucker, S.~Ha, J.~Tan, V.~Kumar,
  H.~Zhu, A.~Gupta, P.~Abbeel, {\em et~al.}, ``Soft actor-critic algorithms and
  applications,'' {\em arXiv preprint arXiv:1812.05905}, 2018.

\bibitem{f3}
T.~Haarnoja, S.~Ha, A.~Zhou, J.~Tan, G.~Tucker, and S.~Levine, ``Learning to
  walk via deep reinforcement learning,'' {\em arXiv preprint
  arXiv:1812.11103}, 2018.

\bibitem{f4}
D.~Silver, G.~Lever, N.~Heess, T.~Degris, D.~Wierstra, and M.~Riedmiller,
  ``Deterministic policy gradient algorithms,'' in {\em International
  conference on machine learning}, pp.~387--395, Pmlr, 2014.

\bibitem{f5}
S.~Fujimoto, H.~Hoof, and D.~Meger, ``Addressing function approximation error
  in actor-critic methods,'' in {\em International conference on machine
  learning}, pp.~1587--1596, PMLR, 2018.

\bibitem{m1}
K.~He, X.~Zhang, S.~Ren, and J.~Sun, ``Deep residual learning for image
  recognition,'' in {\em Proceedings of the IEEE conference on computer vision
  and pattern recognition}, pp.~770--778, 2016.

\bibitem{q8}
A.~Dosovitskiy, G.~Ros, F.~Codevilla, A.~Lopez, and V.~Koltun, ``Carla: An open
  urban driving simulator,'' in {\em Conference on robot learning}, pp.~1--16,
  PMLR, 2017.

\end{thebibliography}

\end{document}